# RESOLVING THE PARADOX OF THE DIRAC EQUATION: PHENOMENOLOGY

S.F. Timashev

*National Research Nuclear University MEPhI,*
*e-mail: serget@mail.ru*

Based on the results of F. Wilf on the need to take into account the quantum-mechanical correspondence rules in the Dirac equation for an electron, it was shown that the equation obtained by giving physical meaning to α-Dirac operators should be considered as a phenomenological equation for a particle of non-zero size – the EM polaron, previously introduced by the author. This allows a solution to be found to the inherent paradox of the Dirac equation, which consists of the equality of the velocity of the moving particles to the speed of light $c$ in a vacuum, which is *a priori* unobtainable, and to understand the physical essence of spin as the intrinsic mechanical moment of an EM polaron. It is also shown that the Dirac–Wilf equation for a single spatial dimension can be considered a generalization of the Schrodinger equation for relativistic energies.

## INTRODUCTION

The relativistic Dirac equation [1] for an electron is one of the basic equations of modern quantum science. It is considered valid for both electrons and other simple particles (opposed to composite particles like nucleons) with spin $s = ½\ \hbar$, where $\hbar$ is Planck's constant (muons and neutrinos in particular). The Dirac equation is also considered to be applicable to quarks (as particles included in the structure of nucleons). There is, however, one confusing item. In accordance with the derivation [2] of this equation, it is valid only for a speed of particles equal to that of light in a vacuum ($c$), which is itself paradoxical. To what extent can we in this case even talk about a quantum mechanical equation, since material particles cannot move at such speeds?

It was shown in [3, 4] that we can solve problems associated with establishing the physical essence of quantum mechanics, including those related to considering effects at relativistic speeds of particle motion by choosing an absolute base frame of reference: the electromagnetic component of the physical vacuum – EM vacuum tied to the expanding (inflating) Euclidean Universe. The scale of global time $t$, which is common for all points of the Universe and is counted from moment $t = 0$ corresponding to the Big Bang, is used. It is also believed that the EM vacuum affects all material bodies of the Universe, specifically electrons and atomic nuclei, which are open systems for the EM vacuum. These effects ae felt through the emergence of Casimir potential energy with certain boundary conditions on the surfaces of electrons and atomic nuclei, along with the creation of regions of Casimir polarization formed by virtual photons in the near-surface region of these particles, which in fact should take the form of nonpoint particles (EM polarons).

The problem facing Einstein when he introduced the gravitational constant into the energy tensor of the General Theory of Relativity (GTR) is indirectly solved by introducing Casimir potentials. Einstein wrote that "instead of the scalar density of matter, we must operate with the energy-momentum tensor per unit volume. The latter includes not only the energy tensor of matter, but also the electromagnetic field. However … the description of matter using the energy tensor, from the viewpoint of a more accurate theory, should be considered only preliminary. In reality, matter consists of electrically charged particles and should itself be considered a part (and moreover the main part) of the electromagnetic field. And only the fact that we do not know enough about the laws of the electromagnetic field of concentrated charges forces us, when presenting the theory, to leave the true form of this tensor still undefined" ([5], p. 68). Einstein's introduction of gravitational constant $G$ as a quantitative parameter of the energy-momentum tensor should therefore be considered a forced (and the simplest) option.

Introducing the Casimir potential energy of particles (electrons and atomic nuclei) in the form of $U(\vec{r}) = -\alpha_C \hbar c/r$, where $\vec{r}$ is the radius vector (we associate the coordinate system with the EM vacuum and assume that a particle at rest is localized at the origin) and $\alpha_C$ is a dimensionless parameter characterizing the intensity of the introduced Casimir interaction ($\alpha_C = \sqrt{2}$; see below) allows us not only to find ways of resolving the paradox inherent in the Dirac equation (see below) but to advance in understanding the phenomenon of gravity as well [3]. By solving the Schrodinger equation in the centrally symmetric field of the Casimir potential, we first obtain an expression for position $\overline{E}_{Vi}$ of the lower energy level that characterizes the bonding energy of the considered particle with the EM vacuum when it is polarized under the influence of this particle, along with corresponding expression $a_{Vi}$ for the size of region EM of Casimir vacuum polarization in the vicinity of particle $i$, i.e., the size of an EM polaron:

$$\overline{E}_{Vi} = -\alpha_C^2 \frac{m_i c^2}{2} \xrightarrow[\alpha_C=\sqrt{2}]{} -m_i c^2; \quad a_{Vi} = \frac{\alpha_C \hbar}{m_i c} \xrightarrow[\alpha_C=\sqrt{2}]{} \frac{2^{1/2}\hbar}{m_i c}. \quad (1)$$

The choice of parameter $\alpha_C = \sqrt{2}$ in (1) to represent the absolute value of $\overline{E}_{Vi}$ (the energy of bonding between a particle and the EM vacuum) in the form of the relation introduced by Einstein for the "rest energy of the particle in question" is quite understandable. The adequacy of this choice of $\alpha_C$ clearly follows from an analysis of energy release $\Delta E$ during radioactive decays, which clearly show the connection between its effects ($\Delta E = \Delta m \cdot c^2$) and change $\Delta m$ in the initial mass of a radioactive substance, determined by the difference between the energies of the initial and final products bonding with the EM vacuum.

It follows from (1) that the radius of the region of Casimir polarization of EM vacuum in the vicinity of an electron with indicated choice $\alpha_C$ is $a_{Ve} = 2^{1/2}\hbar/m_e c$ = 5.2 × 10−11 cm. This value can be taken as the Casimir size of an electron. Let us introduce the value of the characteristic size of the simplest hydrogen atom (its Bohr size $a_B$), defined as the distance of an electron from the nucleus of the hydrogen atom (proton) at which potential energy $U_e(a_B)$ of the former in the field of the latter is equal to the energy of the first level of the discrete spectrum: $U_e(a_B) = -e^2/a_B = -m_e e^4/2\hbar^2$ [6], so that $a_B = 2\hbar^2/m_e e^2 = 1.04 \cdot 10^{-8}$ cm. Since the value of the region of Casimir polarization in the vicinity of the proton is $a_{Vp}$ = 2.82 × $10^{-14}$ cm (i.e., it corresponds to the scale of action of nuclear forces), ratio $a_{Ve}/a_B$ can be considered an indicator of the degree of overlap (interaction) between the region of Casimir polarization of an electron in the ground state of a hydrogen atom and that of its nucleus (a proton). In other words, it is this relationship that connects the dimensionless value of constant fine structure $\alpha_e$ and the dimensionless constant of Casimir interaction:

$$\alpha_e = \frac{a_{Ve}}{a_B}\alpha_C = \frac{1}{137}. \quad (2)$$

It will be shown below that the concepts developed in [3, 4] can be used to solve the problem of the indicated paradoxical nature of the Dirac equation, based on the basic ideas of F. Wilf [7] in applying the quantum mechanical principles of correspondence between operators introduced in constructing this equation to physical characteristics of a moving electron. We start by writing the Dirac equation for a point electron in an external electric field in the Schrodinger representation:

$$i\hbar\frac{\partial\psi(\vec{x},t)}{\partial t}=H_D\psi(\vec{x},t), \qquad (3)$$

$$H_D=c\sum_{j=1}^{3}\hat{\alpha}_j\hat{p}_j+m_0c^2\hat{\alpha}_0+V(\vec{x},t). \qquad (4)$$

Here $H_D$ is the Dirac Hamiltonian; $V(\vec{x},t)$, the potential energy of the electron; $m_0$, the electron rest mass; $\vec{x}=(x_1,\ x_2,\ x_3)$ and *t* are the spatial coordinates and time, respectively; $\hat{p}_j=-i\hbar\partial/\partial x_j$, three operators of momentum components related to ($x_1,\ x_2,\ x_3$); $\psi(\vec{x},t)$, the four-component complex wave function (bispinor). At the same time, Dirac noted that the linear operators $\hat{\alpha}_i$, *i* = 0 to 3, that act on the wave function, should be related to physical characteristics "describing some *new degrees of freedom* related to the *internal motion of the electron*" ([3], Ch. XI, §66), though it is difficult to imagine this internal motion for the electron as a point particle in the Dirac equation. These operators are chosen so that each pair of such operators anticommutes, and the square of each is unity: $\hat{\alpha}_i\hat{\alpha}_j=-\hat{\alpha}_j\hat{\alpha}_i$, where $i\neq j$. Indices *i* and *j* vary from 0 to 3, and $\hat{\alpha}_i^2=1$ for *i* from 0 to 3.

In the considered representation, these operators are expressed by 4 × 4 Dirac alpha matrices:

$$\alpha_0=\begin{pmatrix} I & 0\\ 0 & -I\end{pmatrix},\quad \hat{\vec{\alpha}}=\begin{pmatrix} 0 & \hat{\vec{\sigma}}\\ \hat{\vec{\sigma}} & 0\end{pmatrix};\quad \sigma_1=\begin{pmatrix} 0 & 1\\ 1 & 0\end{pmatrix},\quad \sigma_2=\begin{pmatrix} 0 & -i\\ i & 0\end{pmatrix},\quad \sigma_3=\begin{pmatrix} 1 & 0\\ 0 & -1\end{pmatrix}. \qquad (5)$$

Here, 0 and *I* are 2 × 2 zero and identity matrices, respectively; σ*j* (*j* = 1, 2, 3) represents Pauli matrices introduced for vector spin operator ([8], p. 491).

It must be emphasized that these operators $\hat{\alpha}_i$, were introduced as purely mathematical images without regard to the *correspondence principle* in quantum mechanics. According to this principle, each considered physical characteristic must have a corresponding operator, and vice versa; i.e., each operator in quantum mechanics must have a corresponding physical characteristic, and the formula for transforming the operator of this characteristic must be identical to the formula for the characteristic itself. Dirac noted that operators $\hat{\alpha}_i$ (for *i* from 0 to 3) "describe new degrees of freedom related to some internal motion of the electron" ([2], p. 335). In deriving his equation, however, Dirac was most likely forced to connect the components of the particle velocity operator with those of operator $\hat{\alpha}_i$, rather than those of momentum operator $\hat{p}_j$ as would be expected (see formula (24), §69 from [2]):

$$\hat{\dot{x}}_j=c\hat{\alpha}_j. \qquad (6)$$

As the eigenvalues of operator $\hat{\alpha}_i$ are equal to $\pm1$, relation (6) obviously leads to the fact that when the projection of the velocity of a particle is measured in the presence or absence of a field, the values will be either $\dot{x}_j=+c$ or $\dot{x}_j=-c$. Physically, this result does not make sense. Moreover, as $\lfloor\hat{\alpha}_i,\hat{\alpha}_j\rfloor\neq 0$ at $i\neq j$, when the projection of the velocity of a particle is determined in one direction, it is impossible to simultaneously determine two other velocity projections, which contradicts the experiment. It is quite natural that the paradoxical nature of this conclusion has been noted repeatedly in the literature (see [8, 9] in particular). In connection with the wave equation introduced by Dirac, Pauli wrote in 1933: "In contrast to non-relativistic quantum mechanics, which can be considered logically closed, in relativistic wave mechanics we today have only separate fragments" ([8], p. 529).

As noted in [10], for the physical interpretation of the Dirac equation, it is necessary to consider its other representation, which is different from the original. Specifically, it may be the Foldy-Wouthuysen (FW) representation ([10], Ch.4, *f*), in which the corresponding Hamiltonian $H_{FW}$ is written as $H_{FW}=\hat{\alpha}_0H_D$ and the Dirac equation in the FW representation is valid for any possible, both relativistic and non-relativistic, velocities of the particles under consideration. Despite the lack of understanding of the physical nature of operators $\hat{\alpha}_i$, the effectiveness of using the Dirac equation was clear from its applicability to electrons and other elementary particles with spin $s = \tfrac{1}{2}\hbar$, i.e., fermions (muons, neutrinos) and quarks. As noted in [11], the Sommerfeld–Dirac formula characterizing the fine structure of the spectrum of the hydrogen atom was obtained on this basis, and the value of the Lamb shift, discovered in 1947 (eight years after the Dirac equation was published), was described with an accuracy of around 4%.

DIRAC–WILF EQUATION

The problem of establishing a connection between operators introduced by Dirac (for *i* from 0 to 3) and physical characteristics "*describing some new degrees of freedom related to the internal motion of the electron*" ([2], p. 335) was partially resolved by F. Wilf [7] by considering the internal dynamics of an electron, which determines its spin. Wilf, like Dirac, also considered an electron as a particle of zero volume, but suggested that the trajectory of the electron as a material point would necessarily include the internal dynamics represented by the movement along a sphere with a radius vector $\vec{R}_{e0}$ in which the center acting as the center of inertia of the point particle moves uniformly and rectilinearly in the direction of momentum vector $\vec{p}$. This means that in addition to the directional motion in space, the particle simultaneously rotates about all three coordinate axes passing through the center of the sphere. He assumed that this motion forms the electron spin $s = \tfrac{1}{2}\hbar$, though, admittedly, the above trajectory cannot represent the trajectory of free motion. Since Wilf considered the electron as a point particle, to give his logic physical validity we will follow [3, 4] in considering the electron not as a point particle, but as a particle of finite size – an EM polaron with characteristic size $a_{Ve} = 2^{1/2}\hbar/m_0c$ = 5.2 × $10^{-11}$ cm indicated above. To implement Wilf's procedure, we introduce two dimensional operators in place of $\hat{\alpha}_0$ and $\hat{\vec{\alpha}}$:

$$\hat{\tau}_e = \tau_{e0}\begin{pmatrix} I & 0 \\ 0 & -I \end{pmatrix} \text{ и } \hat{R}_{je} = R_{e0}\hat{\alpha}_0\hat{\alpha}_j = R_{e0}\begin{pmatrix} 0 & \sigma_j \\ -\sigma_j & 0 \end{pmatrix} \equiv R_{e0}\hat{\alpha}_{0j}, \qquad (7)$$

which are associated with a certain time interval $\tau_{e0}$ (period) and components $R_{e0}\sigma_j$ of radius vector $\vec{R}$ of the point of the sphere (the surface of the EM polaron):

$$\tau_{e0} = \frac{2^{1/2}\hbar}{m_0c^2}, \quad R_{e0} = \frac{2^{1/2}\hbar}{m_0c}. \qquad (8)$$

The movement of the center of inertia of the EM polaron is characterized by momentum $\vec{p} = m_u\vec{u}$, where $m_u$ and $\vec{u}$ are the mass and velocity vector of the electron in the base frame of reference associated with the EM vacuum. It should be noted that when introducing parameters $\tau_{e0}$ and $R_{e0}$ of the Dirac equation for a point electron, Wilf assumed that $\tau_{e0} = \hbar/m_0c^2$, $R_{e0} = \hbar/m_0c$.

As for scalar operator $\hat{\tau}_e$, it can be associated with the characteristic time of restructuring of the EM vacuum's region of Casimir polarization in the vicinity of an electron as an EM polaron during its rotation and translational motion, which is accompanied by an exchange of virtual photons with the EM vacuum. A kind of lubrication is obtained during the movement of the particle upon this exchange (see [3]). The Casimir polarization of the EM vacuum in the direction of motion drops sharply at relativistic electron velocities, due to the loss of the lubricant fraction in the front and opposite regions of the electron's Casimir polarization. The resistance to motion increases, and the inertial mass (potential energy) of the electron grows as a result. According to [3], we obtain $m_u = m_0\eta_u \equiv m_0\left(1 - u^2/c^2\right)^{-1/2}$ for the growing ultrarelativistic mass of a moving electron as $u \to c$. The characteristic size of the relativistic polaron $R_{eu} = R_{e0}\left(1 - u^2/c^2\right)^{1/2}$ is reduced sharply in the direction of its movement, due to the loss of the lubricant fraction. Scalar operator $\hat{\tau}_0$ is then associated with shrinking time $\tau_{eu} = \tau_{e0}\left(1 - u^2/c^2\right)^{1/2}$ of restructuring the region of Casimir polarization of the EM of the vacuum, which is necessary for the movement of the electron.

To introduce operators $\hat{\tau}_e$ and $\hat{R}_{ej}$ into Eq. (3), let us follow Wilf [7] by acting on Eq. (3) with operator $\hat{\tau}_e$ and move to the Dirac equation written in the FW representation:

$$i\hbar\hat{\tau}_e \frac{\partial \psi(\vec{x},t)}{\partial t} = \left[ R_e \sum_{j=1}^{3} \hat{\alpha}_{0j}\hat{p}_j + \tau_e m_0 c^2 \hat{\alpha}_0^2 + V(\vec{x},t)\tau_e \hat{\alpha}_0 \right] \psi(\vec{x},t), \qquad (9)$$

Using these operators, the Dirac equation takes the form

$$\left\{ \left[ \hat{\alpha}_0 \left( i\hbar \frac{\partial}{\partial t} - V \right) - m_0 c^2 I \right] - \frac{R_e}{\tau_e} \sum_{j=1}^{3} \alpha_{0j}\hat{p}_j \right\} \psi(\vec{x},t) = 0. \qquad (10)$$

We will call this the Dirac–Wilf wave equation for an EM polaron as a non-point particle with size $R_{e0} = 2^{1/2}\hbar/m_0 c$ and spin ½ħ. Since $R_{e0}/\tau_{e0} = c$ for an electron, Eq. (10) is actually the original Dirac equation, but it is valid for arbitrary particle velocities, nonrelativistic and relativistic. It also becomes clear why the original Dirac equation, which (as indicated) is *a priori* paradoxical, so easily became part of quantum science even when the speed of a particle was notably less than $c$. The author believes this paradox, which was associated with the Dirac equation for more than 90 years, is resolved by considering the electron a Casimir EM polaron with a non-zero size.

F. Wilf [7] associated the physical essence of an electron's own mechanical moment (its spin) with introducing a characteristic size in the rotational motion of an electron. Wilf considered a point electron, however, so he had to introduce a complex trajectory of motion of this material point. This took the form of motion in a circular orbit with radius $R_0$ around an axis passing through the center of the circle, and the general movement of this center in space as the center of mass. Introducing the EM polaron as an object having its own angular momentum and rotating around an axis with a certain angular velocity gives Wilf's idea substance.

To put the nature of the electron spin into concrete form, we can assume that the bulk of a spherical EM polaron is either uniformly distributed or concentrated in its near-surface region. When the former rotates relative to axis passing through the center, its moment of inertia is $J_B = 2/5 \cdot m_0 R_{e0}^2$. When the latter rotates, the corresponding value is $J_S = 2/3 \cdot m_0 R_{e0}^2$. It is natural to assume that frequency $\omega_r$ of the EM polaron's cyclic rotation is estimated as the ratio of linear speed of rotation $u_r$ of the surface to radius $R_{e0}$: $\omega_r = u_r/R_{e0}$. Since the modulus of the intrinsic mechanical moment (the electron spin) is $L_{Se} = \sqrt{s(s+1)}\,\hbar = \sqrt{3}/2 \cdot \hbar$, we can estimate $u_r$ using the relation

$$k_{B(S)} m_0 R_{e0}^2 \omega_r = L_{Se}, \qquad (11)$$

where $k_B = 2/5$ when a uniformly distributed sphere rotates, and $k_S = 2/3$ when a sphere with its bulk concentrated in its near-surface region rotates. After appropriate substitutions, we obtain

$$u_r = \sqrt{3}/2k_{B(S)} \cdot c, \qquad (12)$$

so $u_r = 1.53c$ when the former rotates and $u_r = 0.92\,c$ when the latter rotates.

It should be noted immediately that according to [3], restrictions on speeds below *c* apply only to movements of material objects relative to the basic frame of reference (the EM vacuum) and are associated with ongoing rearrangements of the region of Casimir polarization of particles. Both estimates obtained for a rotating sphere with a uniform distribution are therefore acceptable.

It should be also noted that the above ideas are fully consistent with general views developed earlier on the nature of electron spin [12], according to which it can be interpreted as a circulating flow of energy in the wave field of an electron. As above, a constant feeding of energy is naturally required for the existence of such circulation or the rotation of the region of Casimir polarization, which is possible only from the base medium (the EM vacuum) [4].

To illustrate possible applications of the introduced equation, the next section will consider the simplest examples of using the Dirac–Wilf equation to solve quantum mechanical problems with two dimensions ( $x$ , $t$ ), where an electron is transferred along one coordinate
under the influence of electric field, the potential energy $V(x)$ of which depends only on the coordinate. The problem is in this case simplified, and if we consider the wave function of Eq. (10) to be a bi-spinor defined by four functions $\psi_i(x,t)$, where $i$ = 1, 2, 3 and 4, the Dirac–Wilf (DW) equation breaks down into two identical pairs of two interrelated equations for the wave functions in the form of spinors [13, 14]. Spinor-forming functions $\psi_1(x,t)$ and $\psi_4(x,t)$ are interconnected in one pair, and spinor-forming wave functions $\psi_2(x,t)$ and $\psi_3(x,t)$ are interconnected in the other. The first of these pairs is considered a set of two related equations for the electron wave function; the second pair of equations are the same for the positron wave function (a state with negative energy). It should be emphasized that the resulting pairs of equations are in no way related to each other at a potential energy greater than the energy of the possible formation of an electron-positron pair, where $V(x) > 2m_0c^2$ (see below).

We will limit ourselves here to considering such processes when the generation of particle and antiparticle pairs is excluded. At the same time, we will consider stationary processes when $\psi_i(x,t) = \tilde{\psi}_i(x)\exp(-iEt/\hbar)$, where $E$ is the energy of the system. For the wave function of the electronic subsystem, which is determined by a pair of distinguishable equations for bi-spinor components $\tilde{\psi}_1(x)$ and $\tilde{\psi}_4(x)$ that we subsequently denote as $\tilde{f}(x)$ and $\tilde{g}(x)$, respectively, we obtain the system of equations:

$$
\begin{aligned}
&\left[E - V(x) - m_0c^2\right]\tilde{f}(x) + i\hbar\frac{R_e}{\tau_e}\frac{d\tilde{g}(x)}{dx} = 0,\\
&\left[E - V(x) + m_0c^2\right]\tilde{g}(x) + i\hbar\frac{R_e}{\tau_e}\frac{d\tilde{f}(x)}{dx} = 0.
\end{aligned} \tag{13}
$$

After differentiating the second equation according to the coordinate, this system can be presented as:

$$
\begin{aligned}
&\frac{d^2\tilde{f}(x)}{dx^2} + \frac{1}{E + m_0c^2 - V(x)}\frac{dV(x)}{dx}\frac{d\tilde{f}(x)}{dx} + \frac{\tau_e^2}{\hbar^2R_e^2}\left\{\left[E - V(x)\right]^2 - (m_0c^2)^2\right\}\tilde{f}(x) = 0,\\
&\tilde{g}(x) = -\frac{i\hbar R_e}{\tau_e\left[E + m_0c^2 - V(x)\right]}\frac{d\tilde{f}}{dx}.
\end{aligned} \tag{14}
$$

It should be noted that when the Dirac–Wilf equation is presented in the form (13) and (14), spinor components $\tilde{f}(x)$ and $\tilde{g}(x)$ characterize the wave function and its derivative, so we need only the continuity of functions $\tilde{f}(x)$ and $\tilde{g}(x)$, respectively, to couple solutions at the boundaries of regions with different external potentials.

Note that the replacement of total energy $E$ in the first equation of (14) by the corresponding operator $i\hbar\,\partial/\partial t$ for the total energy of the particle enables us to derive the Klein-Gordon-Fock equation for free non-zero-volume particles of zero spin:

$$
\frac{\partial^2\psi(x,t)}{\partial x^2} - \frac{\tau_e^2}{R_e^2}\left[\frac{\partial^2}{\partial t^2} + \frac{(m_0c^2)^2}{\hbar^2}\right]\psi(x,t) = 0 \tag{15}
$$

Equations (10), (13), (14) and (15) materialize the image of a wave-particle, the basic object of quantum mechanics, which until now has been introduced only hypothetically in the form of a de Broglie wave, which does not fit into the apparatus and equations of quantum mechanics and expresses only the idea of combining wave and corpuscular properties in one object. Due to the compulsory temporal dispersion assigned to it, the usually postulated de Broglie wave-particle should decay at microscopic distances. A serial electric transmission microscope therefore works contrary to the orthodox interpretation, and in electron microscopy we must assume that a de Broglie wave actually accompanies an electron and travels a considerable distance from the cathode to the detector without decaying inside the microscope.

It should be emphasized that introducing the EM vacuum as the basic material medium and the idea of its Casimir polarization in the vicinity of elementary particles and atomic nuclei allowed us to consider the image of a wave-particle within quantum mechanics. The movement of each wave-particle in the EM vacuum as an EM polaron with mass $m_0$, characteristic size $\lambda_{DW} = R_{e0} = \sqrt{2}\,\hbar / m_0 c$ and velocity $u$ actually means the movement of a local heterogeneity of the EM vacuum fixed in size as a solitary wave with the wave number $k_{DW} = 2\pi / \lambda_{DW}$, cyclic frequency $\omega_0 = 2\pi / \tau_0$ of relaxational restructuring of the Casimir polarization domain of EM vacuum near the particle, momentum $p = m_{iner} u$ and energy:

$$E = \sqrt{\frac{R_0^2}{\tau_0^2} p^2 + m_0^2 c^4} = \sqrt{p_{grav}^2 c^2 + m_0^2 c^4} = \sqrt{m_{grav}^2 u^2 c^2 + m_0^2 c^4} = m_{grav}(u) c^2 , \qquad (16)$$

where

$$p_{grav} = m_{grav} u, \quad m_{grav} = \frac{R_0}{\tau_0 c} m_{iner}, \qquad m_{grav} = m_0 \Big/ \sqrt{1 - (u^2 / c^2)} \qquad (17)$$

$m_{grav}$ и $m_{iner}$ – gravitational and inertial masses, respectively ($m_{grav}(0) = m_0$). Here it is taken into account that the expression for the phase velocity $u_{ph}$ of the wave arising in the EM vacuum during the motion of the polaron has the form: $u_{ph} = \omega_0 / k_{DW} = \lambda_{DW} / \tau_0 = R_0 / \tau_0$. In this case, for an electron, in accordance with the Dirac equation, $u_{ph} = c$. For the group velocity $u_g$ of Dirac-Wilf wave as the general solitary wave produced by the EM polaron transferred by the EM vacuum disturbance accompanying this transfer, one can obtain in view of (16) $u_g = dE / dp_{grav} = u.$ As in this case the phase adjustment of the general solitary wave is independent of the frequency and is accounted for by a constant $R_0 / \tau_0$, the time dispersion will be absent and there will be no blurring of the Dirac-Wilf wave-particle for the electron and other stable particles as they propagate.

One can suggest that in view of (17), the modified Dirac-Wilf equation represented by (13) and (14) for the particle transferred along one coordinate together with the Klein-Gordon-Fock equation (15) can be used as a qualitative tool to materialize the de Broglie idea of waves of matter particles produced by the motion of not only elementary particles and atomic nuclei, but also objects of larger scale.

In light of relations (17), the simplest version of the Dirac–Wilf equation considered above for two dimensions ($x$, $t$), where a particle is transferred along one coordinate, can become a quantitative basis for substantiating the idea dating back to de Broglie about waves of matter particles produced during the movement of elementary particles and atomic nuclei, and of large-scale objects as well. A number of works in which the authors studied diffraction and interference caused by flows of molecules of different sizes (from C60 and C70 fullerenes to molecules with molecular masses of ~25,000 amu) have appeared in recent years [15-20]. In these experiments, the waves of matter are the nanometer-sized objects moving at non-relativistic speeds $u$. Analysis of these data including the determination the phase velocity of the recorded signals could enable us to obtain the direct information about the time of relaxation $\tau_0 = \tau_0(u)$ and the effective masses of the transferred molecular objects. The model based on the above equations can make it possible to solve the old problem of calculating the ratio $Q_{g/i} = M_g / M_{iner}$, where $M_g$ and $M_{iner}$ are the gravitational and inertial masses of the object under study. To date, it has been found that the experimental values of $Q_{g/i}$ are equal to unity with an accuracy of $10^{-7} - 10^{-5}$ for light and heavy elements of the periodic system. However, the error rises to $0.5 \cdot 10^{-2}$ for the part of "atom mass that is equivalent to the energy of binding with orbital electrons" {[21], Chap. 14, p. 400}. One can therefore suggest that the analysis of the interferential phenomena produced by the waves of matter, specifically by molecules with various molecular masses, will allow us to qualitatively describe and clearly understand the genesis of the differences between the gravitational and inertial masses.

With respect to the transfer of an electron as an EM Dirac–Wilf polaron, this is entirely consistent with the phenomenon of transmission electron microscopy and the pioneering study of L. Biberman, N. Sushkin, and V. Fabrikant [22], where it was confirmed experimentally that wave properties are characteristic of both individual electrons and flows of electrons. It was shown that even in a low-intensity electron beam where each electron travels through the device independently of the others, the diffraction pattern that appears during long exposures does not differ from those obtained during brief exposures for electron flows millions of times more intense. An electron (kinetic energy = 72 keV) passed through the device in $8.5 \cdot 10^{-9}$ s. The device then remained empty for an average interval 30000 times (!) longer, and only after that did a new electron travel through it. It is obvious that with such a long interval between successive passages, the probability of the simultaneous passage of at least two electrons is negligible. The introduced concepts of the electron as a Casimir polaron allow us to understand the results from this classical but little cited work by V.A. Fabrikant and his students, and to substantiate the basic hypothesis of orthodox quantum mechanics about the adequacy of the image of a Dirac–Wilf wave-particle in relation to elementary particles and atomic nuclei.

## THE DIRAC–WILF EQUATION IN SOLVING SIMPLE QUANTUM-MECHANICAL PROBLEMS

### *"Klein's Paradox"*

When the word "paradox" is mentioned in connection with the relativistic Dirac equation in [1], it is usually not the velocity of particles in the above equation that is meant, but the paradoxical nature of the result in [23], published by Klein in 1929, a year after the publication of [1]. Klein's work (see also [13, 14, 24]) analyzed the possibility of relativistic electrons penetrating through potential barriers, using the Dirac equation to estimate the probabilities of such a process. Quantum mechanics was only three years old at the time. The paradoxical result Klein obtained was that when analyzing the penetration of an electron wave incident on a repulsive, infinitely extended, and fairly high-energy potential barrier $V(x)$ of more than $2m_0c^2$, where $m_0$ is the rest mass of the electron, calculations showed that an electron with a total energy less than that of the potential barrier can tunnel into the region of an infinitely repulsive potential without experiencing the exponential decay characteristic of tunneling. The electron flow reflected from the potential step at electron energies of more than $2m_0c^2$ exceeded the flow incident on it. As subsequent studies showed [25], the one-body problem in this case loses its meaning at electron energies of more than $2m_0c^2$, due to the start of the Schwinger effect [26] – the spontaneous generation of electron–positron pairs from the vacuum in a strong electric field in the region of the growing energy barrier to the electron (the extent of this region should be on the order of the Compton wavelength).

The problem of relativistic electrons scattering on a one-dimensional potential of finite width was considered in [25] using a Klein–Gordon–Fock equation containing a second derivative with respect to time. It was found that the sum of the currents reflected and passing through the barrier was always exactly equal to the current of the incident particles. The excess noted by Klein of the current of particles passing through the barrier over the current of incident particles was due to an increase in the total number of particles as a result of the birth of electron–positron pairs when the action of the barrier's electric field was initiated. As was shown, electrons with energies lower than that of the potential barrier cannot propagate freely in the region of the barrier.

The most general approach to interpreting the Klein paradox on the basis of wave equations that include the second derivative with respect to time was presented in [9], where two types of these equations were analyzed: the Klein–Gordon–Fock equation introduced for scalar material fields, and the quartional equation introduced for spinor fields [27], whose name reflects the general solution to this equation being expressed in terms of four orthogonal bispinors (quartions). It was shown in [9] that when the height of energy barrier $V(x)$ exceeds $2m_0c^2$, the Klein paradox has a clear and consistent interpretation associated with the topology of the space of states of material fields described by quartion equations. The state space of a free particle represents two hyperplanes in four-dimensional space, separated by a band gap with width $E_{gap} = 2m_0c^2$. One of the hyperplanes is associated with positive frequency solutions to wave equations; the second, to ones with negative frequency. If the height of a potential energy jump is less than the band gap ( $V(x) < 2m_0c^2$), particles whose states belong to positive and negative frequency bands evolve independently of one another in a stationary case.

This analysis completely confirmed the main conclusions in [25]; i.e., it eliminated virtually all questions related to the Klein paradox. However, transitioning from the Dirac wave equation using a firstorder time derivative to the Klein–Gordon–Fock wave equation and quartion equations that include a second time derivative actually alters the originally considered problem, which is oriented toward using the Dirac equation. The dimension of the relativistic state space of the Klein–Gordon–Fock equation is twice wider than that of the state space of the material field, which is described by equations with the first time derivative, as when considering the Schrodinger equation. With equations that include the second time derivative, we can therefore arbitrarily specify wave function $\psi$ and $\partial\psi/\partial t$ for a certain moment of time $t$. In contrast to the Schrodinger equation, which allows a probabilistic interpretation of the considered processes, the expression for particle density $\rho_0 = \psi\psi^*$ is generally not a positive definite quantity. It also indicates the need to generally consider particles with different charge signs (electrons and positrons) simultaneously [11].

Namely, the postulation of the generation of electron–positron pairs according to the Schwinger mechanism in a strong electric field is usually meant when discussing the Klein paradox. At the same time, assessing the critical value of electric field strength $F_{cr}$, above which such pairs can be produced in a constant electric field in a vacuum ($F_{cr} = \pi m_0^2 c^3 / e\hbar \approx 4 \cdot 10^{16}\, V/cm$; see (6.41) in [26]), shows that such strengths are really unobtainable. The formation of electrons and positrons recently discovered in [27] under conditions of applying a strong electric field to a system formed by a graphene superlattice on hexagonal boron nitride, and interpreted by the authors as a mixture of Zener and Klein tunneling, can hardly be associated with the process predicted by Schwinger. The author believes interband tunneling may have occurred in this system; i.e., transitions of electrons, initiated by a repulsive electric field from negative levels in a system of filled energy bands (a "Dirac sea" [28]) created by electronic subsystems of real atoms in the system studied experimentally in [27], to positive levels and the formation of holes as positrons in filled states, in analogy with known processes in semiconductors.

However, the question remains as to what extent the ejection considered by Schwinger of a particle and an antiparticle from a vacuum can be realized when a sufficiently strong electromagnetic field is applied in the absence of any other particles. According to the concepts developed in [3, 4], the electron is a nonpoint particle due to the polarization of the EM vacuum in its vicinity and thus the formation of a region of Casimir polarization created by virtual photons and having characteristic size $a_{Ve} = 2^{1/2}\hbar / m_e c$ = 5.2 × $10^{-11}$ cm. It is important that localized virtual photons at given wavelength $\lambda$ and energy $\hbar\omega/2$ are associated with mass $\Delta m_\lambda = \hbar\left(2\pi c\lambda^{-1} - \omega_{eff}\right)/2c^2$, where frequency $\omega_{eff} = 2\pi u_{eff}/\lambda$ is determined by effective speed of light $u_{\text{eff}}$ in the region of Casimir polarization of the EM vacuum in the vicinity of an electron. We then have upon complete localization of a virtual photon when $u_{eff} \to 0$, so $\Delta m_\lambda = \hbar\omega/2c^2$. This means an induced dipole component and seed mass inevitably appear in localized virtual photons due to the influence of the Coulomb field of the initial electron, and they both should increase when exposed to an external electric field. The question of how much strength $F_{cr}$ of the electric field can fall relative to the above value "according to Schwinger" as a result of such processes, so that the heterolytic dissociation of virtual photons occurs with the formation of $e^+e^-$ pairs in the EM vacuum with an electron acting as a catalyst, remains open. From this viewpoint, experimental studies of the possibility of $e^+e^-$ pairs forming when photons with energies $E_{ph} \le 2m_0c^2$ act on electron flows in strong electric fields (a kind of analog of the Keldysh–Franz effect [28]) could be of interest. Since all fundamental questions related to the problem of the Klein paradox can be considered resolved, it might be interesting to obtain specific expressions for the electron wave function in the Klein problem when considering the Dirac–Wilf equation for arbitrary electron energies outside energy barrier $V(x)$ and in the barrier region. Such formulation of the problem is excluded when using the original relativistic Dirac equation. To be definite, we limit ourselves here to considering when $V(x) < 2m_0c^2$ so that no electron–positron pairs are formed. We will assume that an electron wave from a one-dimensional region $(-\infty < x < 0)$ falls on a repulsive, infinitely extended (region $0 \le x < \infty$) potential barrier $V(x)$. We will also present total energy of the electron $E$ in the form $E = \varepsilon + m_0c^2$, where $\varepsilon$ is kinetic energy of an electron for which $\varepsilon < m_0c^2$.

Representing electrons incident on the potential barrier from region $x < 0$ at $V(x) = 0$ as plane waves with momentum $p = m_u u$ (i.e., assuming that $\tilde{f}(x) = f(x)\exp(ipx/\hbar)$ and $\tilde{g}(x) = g(x)\exp(ipx/\hbar)$) we obtain a system of equations for *f*(*x*) and *g*(*x*) based on (13):

$$\begin{aligned} &\left[E - V(x)\text{-}m_0c^2\right]f(x) - pcg(x) = 0, \\ &\left[E - V(x) + m_0c^2\right]g(x) - pcf(x) = 0. \end{aligned} \tag{18}$$

From the condition that the determinant of this system of equations be equal to zero, we obtain the relation

$$p^2c^2 = E^2 - \left(m_0c^2\right)^2, \tag{19}$$

with which we determine the wave function in region $(-\infty < x < 0)$ for incident and reflected waves from Eqs. (18) when introducing coefficient of reflection *B*:

$$\psi(x,t) = \left\{\begin{pmatrix} p/\hbar \\ \left(E - m_0c^2\right)/\hbar c \end{pmatrix}\exp\left(\frac{i}{\hbar}px\right) + B\begin{pmatrix} -p/\hbar \\ \left(E - m_0c^2\right)/\hbar c \end{pmatrix}\exp\left(-\frac{i}{\hbar}px\right)\right\}\exp\left(-\frac{i}{\hbar}Et\right) \tag{20}$$

We believe that components $\tilde{f}(x)$ and $\tilde{g}(x)$ are proportional to $\exp(-ipx/\hbar)$ for the reflected wave.

Using relations $\exp(-ipx/\hbar)$ and $p=\sqrt{2m_0\varepsilon\left(1+\frac{\varepsilon}{2m_0c^2}\right)}$ at $\varepsilon/m_0c^2<1$, the expression for the electron wave function in the region of the repulsive barrier ($0\le x<\infty$) can be obtained from expression (19) with substitutions $E\to E-V$ and $p\to iq$, where

$$p=\sqrt{\frac{(V-\varepsilon)^2}{c^2}-2m_0(V-\varepsilon)}\,,\qquad q=\sqrt{2m_0(V-\varepsilon)-\frac{(V-\varepsilon)^2}{c^2}}\,. \tag{21}$$

By introducing coefficient $F$ of a wave penetrating into the region of the repulsive potential, we therefore obtain

$$\psi(x,t)=F\begin{pmatrix}-\frac{i}{\hbar}q\\ \frac{V\text{-}\varepsilon}{\hbar c}\end{pmatrix}\exp\left[-\frac{1}{\hbar}qx-i\frac{1}{\hbar}Et\right]. \tag{22}$$

for the wave function in the region $0\le x<\infty$.

The above expressions for the wave functions show that the Dirac–Wilf equation can be used adequately over the range of possible non-relativistic and relativistic electron energies when $\varepsilon/m_0c^2<1$. At high electron energies, we must consider the generation of $e^+e^-$ pairs at high strengths of the electric field in the region of the potential barrier boundary.

As boundary conditions, let us consider the continuity of wave functions (20) and (22) at boundary $x=0$ [12, 13]. The corresponding equations for determining coefficients of reflection and penetration $B$ and $F$ for a wave in the region of the repulsive potential have the form

$$\begin{aligned}&p(1-B)=-iqF,\\&\varepsilon(1+B)=(V-\varepsilon)F.\end{aligned} \tag{23}$$

Since $q/p=i,$ it follows from (23) that

$$B=1-\frac{2\varepsilon}{V},\quad F=\frac{2\varepsilon}{V},\quad . \tag{24}$$

so

$$B+F=1. \tag{25}$$

This means that an electron as a wave-particle at an energy less than that of the considered barrier is partially absorbed by it and transferred into the depth of the barrier. As follows from introduced relativistic corrections (21) to tunneling momentum $q$ under adopted restrictions $2\varepsilon/V<1$ when $V/2m_0c^2<1$, the depth of penetration into the barrier grows substantially for ultrarelativistic electrons.

*Rectangular Potential Box*

As a second example of using the Dirac–Wilf equation to analyze the one-dimensional motion of an electron, let us consider the motion in a rectangular potential well [6] representing the coordinate dependence of electron potential energy $V(x)$ in the form

$$\begin{cases}V(x)=V_0<0 & \text{at } 0<x<a,\\ V(x)=0 & \text{at } x<0,\ x>a.\end{cases} \tag{26}$$

It is obvious that under the condition $V_0<E<0$, where $E$ is the electron energy, the spectrum of the electron must be discrete; if $E>0$, there is a continuous spectrum of doubly degenerate levels. Due to (8), it follows from Dirac–Wilf Eq. (14) that in region $0<x<a$,

$$\frac{d^2\widetilde{f}(x)}{dx^2}+\frac{\tau_e^2}{\hbar^2 R_e^2}\left[(E-V_0)^2-\left(m_0c^2\right)^2\right]\widetilde{f}(x)=0,$$

$$\widetilde{g}(x)=-\frac{i\hbar R_e}{\tau_e\left[E+m_0c^2-V(x)\right]}\frac{d\widetilde{f}}{dx}. \tag{27}$$

By only illustrating the determination of electron energy levels $E$ in such a potential well, we limit ourselves to considering the limit case (see [6], § 22) of fairly high walls of potential well $V_0$, where the electron moves only in the area limited by points $x=0$ and $x=a$ so that exiting the tunnel beyond this section is excluded, and at the indicated points conditions and nust be selected for wave functions $\widetilde{f}(x)$ and $\widetilde{g}(x)$.

We seek the solution to $\widetilde{f}(x)$ in the form

$$\widetilde{f}(x)=C\sin\left(\frac{p}{\hbar}x+\delta\right), \quad . \tag{28}$$

where

$$p=\frac{\tau_e}{R_e}\sqrt{(E-V_0)^2-\left(m_0c^2\right)^2}\,. \tag{29}$$

From condition $\widetilde{f}=0$ at $x=0$, it follows that $\delta=0$, after which we obtain equality $\sin(pa/\hbar)=0,$ from the same condition at $x=a$, and

$$pa/\hbar=n\pi\,, \tag{30}$$

Where $n$ are positive integers starting from one. After representing total energy $E$ of an electron in the form $E=E^*+m_0c^2$ and substituting this expression into (29), in light of (8) and based on Eq. (30) we obtain a quadratic equation for the positions of electron energy levels $\varepsilon_n$ relative to the bottom of the considered potential well:

$$\left(E_n^*-V_0\right)^2+2m_0c^2\left(E_n^*-V_0\right)-\frac{\pi^2\hbar^2c^2}{a^2}n^2=0\,. \tag{31}$$

Therefore,

$$E_n^*-V_0=m_0c^2\sqrt{1+\frac{\pi^2\hbar^2c^2}{\left(m_0c^2\right)^2a^2}n^2}-\left(m_0c^2\right)^2\approx\varepsilon_n\left[1-\frac{\varepsilon_n}{2m_0c^2}+\frac{2\varepsilon_n^2}{\left(2m_0c^2\right)^2}\right], \tag{32}$$

where the first term on the right side of reducible expansion

$$\varepsilon_n=\frac{\pi^2\hbar^2}{2m_0a^2}n^2 \tag{33}$$

is the expression considered in nonrelativistic quantum mechanics to determine the energy levels of a particle in a rectangular potential well [6]. It should be noted that the given expansion, showing if $2\varepsilon_n^2/\left(2m_0c^2\right)^2<<1$, relativistic corrections to $E_n$ are valid up to fairly high levels of excitation where $\varepsilon_n$ can be comparable to $m_0c^2$ (the second term on the right side of (32)).

In accordance with (32), the wave function (components of bi-spinor $\widetilde{f}(x)$, $\widetilde{g}_n(x)$) and momentum $p$, which determines wave function (28) for each possible values of energy $\varepsilon_n$, should naturally be characterized by subscript (i.e., presented as $\widetilde{f}_n(x)$, $\widetilde{g}_n(x)$, and $p_n$, respectively). Expressions for the normalized components of bi-spinor $\widetilde{f}_n(x)$ and $\widetilde{g}_n(x)$ have the forms:

$$\widetilde{f}_n(x)=\sqrt{\frac{\varepsilon_n-V_0+2m_0c^2}{a(\varepsilon_n-V_0+m_0c^2)}}\sin\left(\frac{p_n}{\hbar}x\right),\quad \widetilde{g}_n(x)=-i\sqrt{\frac{\varepsilon_n-V_0}{a(\varepsilon_n-V_0+m_0c^2)}}\cos\left(\frac{p_n}{\hbar}x\right). \tag{34}$$

The simple examples considered in this section give reason to believe that Eqs. (13) and (14) can be considered a generalization of the one-dimensional Schrodinger equation to the case of arbitrary (including relativistic) speeds of particles of non-zero size.

## CONCLUSIONS

We have established why the Dirac equation, despite its apparent paradox (the speed of particles is equal to the speed of light in vacuum!), complies fully with experimental data and is reasonably considered the basic equation of modern quantum science. Understanding of these reasons became possible thanks to the work of Wilf [7], who, in search of correspondence of the Dirac equation to the canons of quantum mechanics, specifically, the correspondence principle, according to which an observable physical quantity should correspond to each operator, "saw" the possibility of introducing into the Dirac equation on the basis of the $\alpha$ -matrices-operators of the inherent electron internal structure – its own the radius vector and some characteristic time. It should be immediately noted that when constructing his equation, Dirac understood that his $\alpha$ -matrices must be associated with physical characteristics describing *new degrees of freedom related to the internal motion of the electron*, but it was impossible to do this within the concept of an electron as a point particle. Wilf, who also considered the electron a point particle and relied on new operators he introduced that characterized the dynamics of the electron itself while showing it had its own mechanical moment of $\frac{1}{2}\hbar$ -spin, proposed a variant of the possible trajectory of the electron which differed from that of a freely moving electron.

However, only the use of ideas developed earlier by the author for the electron as an EM polaron of finite size [3, 4] allowed us to generalize the Dirac equation and present the Dirac–Wilf equation for the electron as one for a particle of finite size at arbitrary energies, non-relativistic and relativistic. A 90-year-old problem was thus solved: the reason for the paradoxicality of the relativistic Dirac equation was established, and the physical essence of the emergence of spin as a mechanical moment of the electron was understood. It was also shown that the Dirac–Wilf equation for one spatial dimension, where wave functions are introduced in the form of spinors rather than bi-spinors, can be considered a generalization of the Schrodinger equation to the case of relativistic energies.